\documentclass[aps,epsfig,floatfix,twocolumn,nofootinbib,showpacs]{revtex4}
\usepackage{graphicx}
\usepackage{latexsym}

\begin{document}

\title{Communication Bottlenecks in Scale-Free Networks}

\author{Sameet Sreenivasan,$^{1,2}$ Reuven Cohen,$^{3}$ Eduardo L\'{o}pez,$^4$\\
Zolt\'{a}n Toroczkai,$^2$\footnote{Department of Physics, University of
  Notre Dame, Notre Dame, IN 46556 (after Jun. 1 2006)} and H. Eugene Stanley$^{1}$ }

\affiliation{$^1$Center for Polymer Studies and Department of Physics, Boston
  University, Boston, MA 02215\\
$^2$ Center for Nonlinear Studies, Los Alamos National Laboratory, MS B258, Los 
Alamos, NM 87545\\
$^3$ Laboratory of Networking and Information Systems and Department of
Electrical and Computer Engineering, Boston University, Boston, MA 02215\\
$^4$ Theoretical Division, Los Alamos National Laboratory, MS B258, Los
Alamos, NM 87545
 }

\date{\today}

\begin{abstract} We consider the effects of network topology on the
optimality of  packet routing quantified by $\gamma_c$, the rate of packet
insertion beyond which congestion and queue growth occurs.  The key result of
this paper is to show that for any network, there exists an absolute upper
bound, expressed in terms of vertex separators, for the scaling of $\gamma_c$ with
network size $N$, irrespective of the routing algorithm used. We then
derive an estimate to this upper bound for scale-free networks, and introduce
a novel static  routing protocol which is superior to shortest path routing
under intense packet insertion rates.
\end{abstract}

\pacs{89.75.Hc, 89.20.Hh, 89.75.Da }
\maketitle

Communication has stepped into a new era with the advent of the Internet, making
possible information exchange/transport across the globe virtually in an instantaneous
fashion between any two people who have access to it. Broadcasting and
advertising messages, home-pages, blogs and practically any information posted on 
the WWW is within the reach of anyone accessing those pages and thus,
downloading that information. This activity, exponentially increasing over the past years 
involves an incredible amount of information  stored and transmitted through the physical 
infrastructure of the Internet, every second of the day. As the number of computers and users
surpasses into the billions, one might naturally ask about the ultimate limits to using the
Internet. In terms of {\em transmission latency} the Internet is pretty good already.
As an illustration, consider the distance between Los Alamos and Boston (as the crow flies),
which is about $3109\;km$. The speed of light in fiber is about 2/3 of that in vacuum, about
$2\times 10^5 \;km/s$. Thus the round-trip time for information between Los Alamos and
Boston is about $31\;ms$. Performing a ping on a Los Alamos computer to a computer
at Boston University gives for the round-trip time about $64\;ms$ which is within a 
factor of two of the absolute physical bound. Therefore, no order of magnitude 
improvements can be  expected in transmission latency for the Internet. 
The current paradigm in communication on networks is packet switching where
the message is divided into packets which are then 
routed between nodes over data links, independently from each other, and reassembled 
at the destination into the original message. 
This decentralized methodology makes information transmission efficient by providing better 
utilization of the available bandwidth (a single link can be used 
to transmit any packet). 
However, due to the increasing demand of information carried through
the Internet, delays can occur in packet delivery, mainly caused by device (end-user and 
router) latency. Device latency is the amount of time $\tau$ that a device needs 
to process a single packet. Although the devices are getting better in their latency, this is a physical
constraint and can never be completely eliminated.  Since more packets may arrive at a 
node than it is able to process per unit time, queues can accumulate and thus routers must  
have a storing  capacity as well.  These queues will naturally slow down information 
transport over the network.
As an interesting observation, the US Postal Service is capable of achieving
higher information transmission rates than the current Internet. For example, for a T1 line 
which transmits at about 1.544 Mbit/s, downloading a 4.7 GB DVD takes about 6.76hrs. If
one ships 1000 DVD-s from coast to coast in the US, it will take about 3 days, 
but the transmitted information would have a bandwidth matching that of 94 T1 lines. 
Precisely this fact is exploited by DVD rental delivery companies like Netflix which distributes 
about 1.5 Terabytes of data per day, the same order of magnitude as the
Internet \cite{NYT02}. 

In spite of technological advances, the Internet is being driven closer to its capacity. 
These facts lead us to two important questions: (1) How can one characterize a packet switched
communication network's ultimate carrying capacity? and (2) What routing algorithms will achieve this
ultimate capacity?

In this Letter we present a proof-of-principle study to show that the ultimate carrying capacity is 
strongly influenced by the network's structure. We demonstrate the existence of a solely topology determined 
upper bound $\gamma_T$ for the congestion threshold $\gamma_c$ \cite{Fuks99} which is the packet insertion 
rate at which queuing and congestion in the network appears.

It has been conjectured that the degree distribution of the Internet follows a power 
law on several levels \cite{Fal99,Tang02,Vazquez}. Recent 
experimental studies have strengthened the validity of this conjecture 
\cite{CAIDA,DIMES}. For our study, we will confine ourselves to the the 
configuration  model (CM) \cite{Molloy95}  which is one of the simplest models to
generate a random graph with a power law degree distribution. The approach 
presented here is, however, applicable to arbitrary graph  structures. 

 We consider all time scales measured in units of 
router latency $\tau$ which for  simplicity, we take to be unity.  We will 
also assume that routers have infinite storage capacity.

{\em The Static Routing Problem.}   
Denote by $G(V,E)$ the physical substrate graph (network) for communication which we
assume to be singly connected. 
Once a packet entering node $s$ reaches its destination node $d$, it disappears 
from the system. The sequence of nodes and edges the packet visits constitutes
the route for that source-destination pair. For a network of size $N$, the routing problem consists of finding an assignment of 
routes for {\em all}  $N(N-1)/2$ pairs of nodes.  We shall 
call such an assignment set a {\em Static Routing Protocol} (SRP). 
%We restrict
%ourselves to  {\em loopless} routes, i.e., a node in the route appears only once
%in the same route. 

We consider a previously studied \cite{Fuks99,Arenas01,Sole01,Zhao05} model
of communication, which was motivated by the need to study the problem of
congestion on the router-level Internet. Here, the packet transmission is
modeled by a discrete time parallel update algorithm.  At time $t$ and at every node, a packet enters with probability $0
\le \gamma \le 1$. The packet has a destination node, chosen uniformly at
random from the remaining $N-1$ nodes. Every node $i$ maintains a set of all 
packets that were sent to it by its neighbors in the previous step,
eliminates from this set newly arrived  packets whose destination was $i$,
adds to this set the freshly injected packet (if there is one) and finally
places elements of this set in a sub-queue in a random order. This randomization is needed
because times are not resolved below the single-packet processing timescale,
$\tau$. The sub-queue is then appended to the existing queue, if there is one, from
before the $t$-th step. The top packet in the queue is then sent to a neighbor
on $G$ following the SRP.

There is a critical rate
$\gamma_c$ of packet creation at which there is an onset of congestion,
i.e., above $\gamma_c$, packets start accumulating on the network \cite{Arenas01,Zhao05}. 
This is commonly designated as the ``congestion threshold''. In Fig.~\ref{fig:3}, 
we show a rescaled version \cite{Arenas01} of the rate of steady-state packet growth  
$\theta(\gamma) \equiv \lim_{t\to \infty} \left[ n(t+\Delta t)- n(t)\right]/\left( N\gamma \Delta t\right)$
as function of $\gamma$ for both the shortest path (SP) protocol
and the novel one proposed in this paper. 
Here $n(t)$ is the number of packets on the network at time $t$.
This threshold can be expressed in terms of the maximal node betweenness
$B$ for a given SRP. The betweenness $b$ for a node is the number of SRP
routes passing through that node. The highest among the $N$ betweenness values
(one for each node) resulting from the SRP is the maximal
node betweenness $B$. For a given SRP route
between a source $s$ and destination $d$ the average packet current incurred from 
the source at $s$ is $\gamma/(N-1)$. 
For a node with betweenness $b$ the average packet inflow current will be given by 
$b \gamma/(N-1)$. Since the outflow of packets occurs at unit latency, 
we will have queueing and congestion at the node for which this quantity reaches unity
for the first time, namely at the node with $b=B$. Thus
\begin{equation}
\gamma_c =\frac{N-1}{B}. \label{E.1}
\end{equation}
For SP routing 
\cite{Fuks99,Arenas01,Sole01,Zhao05}, the node betweenness becomes
identical to the familiar, shortest path betweenness, $B^{SP}$
\cite{Freeman77}.  From Eq.~(\ref{E.1})  follows that for a given routing
protocol, the dependence of the congestion threshold $\gamma_c$ on $N$, is
determined by the scaling with $N$ of the maximal node betweenness
$B$. Therefore, the best routing protocol from the
point of view of router congestion avoidance, should be the one for which
$B$ exhibits the {\em slowest} growth with $N$. Although there have
been prescribed {\it ad hoc adaptive} protocols
\cite{Ech04,Ech05,Yan05} that increase $\gamma_c$, 
the above issue has not been systematically addressed. 

\begin{figure} 
\includegraphics[width=7.5cm,height=1.5cm,angle=0]{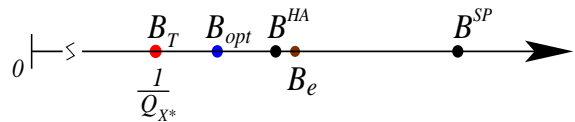}
\caption{The relative sizes of the betweenness values introduced in the text.}
\label{fig:bsaxis}
\end{figure}

Next, we show that there is a lower bound $B_T \leq B$ 
(and thus $\gamma \leq \gamma_T$) induced 
only by the {\em topology} of the network $G$, and it is independent of the routing
protocol used. In other words, no SRP can do better than $\gamma_T$. This
ultimate threshold $B_T$ is essentially a communication bottleneck quantifier for 
a given graph $G$. 
Among all possible SRPs (whose set is  denoted as ${\cal P}$), let us write 
$B_{opt}$ for the smallest maximal betweenness value, namely 
$B_{opt} = \min_{SRP \in {\cal P}} B^{SRP}$, so  
$B_{T} \leq B_{opt}$ (Fig. \ref{fig:bsaxis}). It is 
an open question whether the topological
bound can be achieved by a routing protocol.
Similar considerations 
have been made in the context of edge betweenness in Refs.\cite{Gkan03, Akella03}.
Here we focus on scaling of the bound $B_T$ as function of $N$. 

We introduce $B_T$ using graph partitioning arguments. 
Given an arbitrary network $G$, partition the set of all nodes $V$ into three non-empty sets denoted, $A$, 
$X$ and $B$. Since $G$ is singly connected, there will be edges running between 
at least two pairs of the three possible pairs. Choose set $X$ such that there are no edges running
directly between $A$ and $B$ in which case $X$ is called a
 {\em vertex separator}.  
 For any SRP we must designate a route for all
 pairs of nodes, therefore also for those pairs for which one node is in $A$
 and the other in $B$.  
 Since $X$ is a separator set, all routes from $A$ to $B$, must go through the nodes
in $X$. Therefore, there are at least $|A| |B|$ routes passing through $X$ for any SRP.  
Since the maximum is always larger or equal than the average, the maximum betweenness
incurred on the nodes in $X$ can be no less than $\frac{|A| |B|}{
|X|}$. We define the {\em sparsity}  \cite{Vazirani01} of the separator $X$ the quantity
$Q_X \equiv \frac{|X|}{|A| |B|}$.  Thus, associated with every vertex
separator $X$ there is a quantity $B_X = 1/Q_X$ providing a lower bound to the
maximal betweenness on nodes in $X$. Let us denote by ${\cal M}$ the set
of all possible vertex separators in $G$. If we systematically consider all
possible choices of vertex separators $X \in {\cal M}$, we can find (at least) one separator
$X^*$ for which $B_{X} =1/Q_{X}$ achieves its maximal value defined as 
$B_T$. Thus, the {\em topology} of the graph constrains the maximal
betweenness to be no less than $B_T$, and for arbitrary routing,
$B \geq B_T = 1/Q_{X^*}=1/\min_{X\in {\cal M}} Q_X$.  Finding minimal 
sparsity vertex separators is an NP-hard problem \cite{Bui92}, and 
we shall not deal with it here.   

Due to the analytical and the computational difficulty in determining
$B_T$, we focus on obtaining an {\em analytical estimate} $B_e$ to $B_T$, and derive
its scaling with $N$ for random,
uncorrelated, scale-free networks.   
This estimate, while possibly being greater than the true topological bound
$B_{T}$, nevertheless provides a comparative value dependent only on the
network topology. This estimate, $B_e$, allows us to quantify the
performance of the SP protocol. 

%We will show that  \begin{equation}
%B_{e} = O(N^{\frac{\lambda}{\lambda-1}}) \label{E.3}
%\end{equation}
%for a random uncorrelated scale-free network of size $N$, degree exponent 
%$\lambda$, and minimal degree $k_{min}\geq 3$. Next, we
%compare the numerically obtained scaling of the maximal betweenness $B^{SP}$
%of Shortest Path routing with the scaling of $B_{e}$, and show that $B^{SP}$
%scales much faster with $N$. This suggests that a SRP for which the maximal
%betwenness scales like $B_{e}$ would have a better performance than SP
%from the point of view of congestion. The question arises whether $B_e$ can be
%achieved by {\em any}  static routing protocol. We answer this question affirmatively by 
%presenting the {\em hub avoidance} (HA)  protocol, and show that it is
%superior even to $B_e$ ($B^{(HA)} < B_e$) and  much better than  the shortest 
%path (SP) protocol. 

\begin{figure}
\includegraphics[width=6.5 cm, height=3.0cm,angle=0]{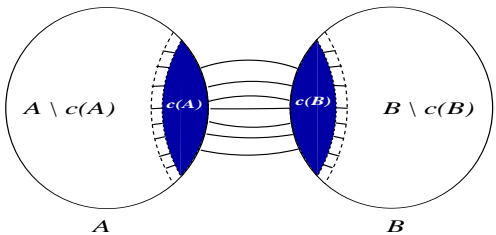}
\caption{Bipartitioning the graph into two vertex subsets $A$ and $B$ such as to obtain
two vertex separators, $c(A)$ and $c(B)$, see text.} \label{fig:css}
\end{figure}

\begin{figure}
\includegraphics[width=4.5 cm,height=7.0cm,angle=270]{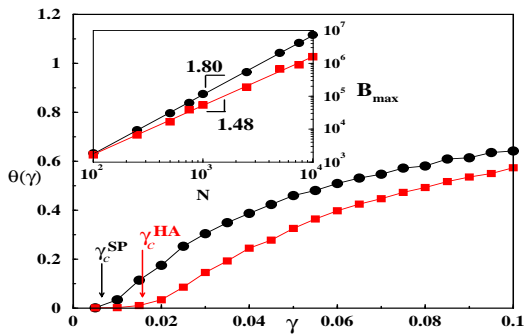}
\caption{Numerical comparison for the performance of SP 
and hub avoidance HA protocols on a scale-free graph of size $N = 10^3$ and
$\lambda = 2.5$. The black circles correspond to the SP protocol and the red
squares correspond to the HA protocol. The congestion threshold $\gamma_c$
beyond which packet growth occurs ($\theta(\gamma) > 0$), is higher for the
HA protocol as compared to the SP protocol. The inset shows that maximum betweenness
for SP and HA protocols on scale-free graphs has power-law scaling with system size.  
The maximal betweenness  $B^{HA}$ resulting from the HA protocol has scaling
exponent $1.48$, close to our estimate for
the topological bound on the maximal betweenness $B_e \sim
N^{\frac{3}{2}}$. However, the maximal betweenness  $B^{SP}$
resulting from the SP protocol grows much faster, $B^{SP}\sim N^{1.80}$. 
} \label{fig:3}
\end{figure}

We start by systematically considering every possible vertex separator in the
graph as follows. First, bipartition the graph as
shown in Fig.~\ref{fig:css} into sets $A$ and $B$ with $|A|\leq |B|$. 
Let $c(A)$ be the subset of nodes in $A$
which are adjacent to at least one node in $B$ and let $c(B)$ be
the subset of nodes in $B$ which are adjacent to at least one
node in $A$. We can now obtain a vertex separator $c(A)$ which separates
sets $A \setminus c(A)$ and $B$, or similarly, a vertex separator $c(B)$ which separates
sets $B \setminus c(B)$ and $A$ .  Thus, going
through all possible bipartitions of the graph with $|A| \leq N/2$
ensures that we have considered all possible vertex separators of the graph. 
If $c(A)$ is chosen as the separator then the sparsity is
$Q_{c(A)} = |c(A)| /  \left( |A - c(A)| |B| \right) 
\geq |c(A)| / \left( |A|   |B| \right)$. We obtain a similar expression for
$Q_{c(B)}$ if  $c(B)$ is chosen as the vertex separator. Therefore
\begin{equation}
Q_{c(A)} \geq  \frac{1}{|B|} \frac{|c(A)|}{|A|}\;\;\;
\mbox{and}\;\;\; Q_{c(B)} \geq  \frac{1}{|B|} \frac{|c(B)|}{|A|}.  \label{E.4}
\end{equation}
Since $|A| \le N/2$, $|B| \equiv O(N)$, and a lower bound for
the sparsity $Q_{X^*}$ is determined by 
\begin{equation}
Q_{X^*} \geq \frac{1}{O(N)}\min_{A\subset V,A \le \frac{N}{2}} \left
  \{\min \left(\frac{
    |c(A)|}{|A|},\frac{| c(B)|}{|A|}\right)\right \}. \label{E.6}
\end{equation}
Next we use the notion of {\it edge expansion} $\chi_e$ defined below.  
For a bipartition of the graph $G$ into sets $A$
and $B$, denote the number of edges simultaneously adjacent to a
node in $A$ and $B$ as $c_e(A,B)$. Then
\begin{equation}
\chi_e = \min_{A \subset V,A \le \frac{N}{2}} \frac{| c_e(A,B)|}{| A |}, 
\label{E.7}
\end{equation}
and an {\it edge expander} graph has $\chi_e \ge O(1)$.
Next consider a bipartition of the graph into $A$ and $B$, and let
$|A| = c N^{\alpha}$ where $c$ is a constant and $0 < \alpha \le
1$. From the edge expansion property of scale-free graphs with $k_{min} \geq 3$ 
\cite{Gkan03}, the number of cut
edges between $A$ and $B$ is at least $\chi_e c N^{\alpha} = O(N^{\alpha})$. 
We can bound from below both $|c(A)| $ and $|c(B)|$ (as needed by
(\ref{E.6})) by the minimal size $m$ of the set of nodes that can
contribute $\chi_e c N^{\alpha}$ cut edges. The size $m$ is obtained by
taking all nodes with degree higher than $\hat{k}$, such that
$N\int_{\hat{k}}^\infty kP(k){\rm d}k=\chi_e cN^\alpha$, 
where $P(k)= A k^{-\lambda}$ is the degree distribution of the graph. This yields
$\hat{k} \sim N^\frac{1-\alpha}{\lambda-2}$.
Therefore the minimal size of the set of nodes that can contribute 
$\chi_e c N^{\alpha}$ edges is:
$m = N\int_{\hat{k}}^\infty P(k){d}k\sim
N\cdot N^{(1-\lambda)\frac{1-\alpha}{\lambda-2}}$
and therefore,
\begin{equation}
|c(A)|, |c(B)| \ge m 
= O\left(N\cdot N^{(1-\lambda)\frac{1-\alpha}{\lambda-2}}\right). \label{cab}
\end{equation}
The quantity $m$ is bounded below by $O(1)$. For a given $\lambda
$ we see that when $\alpha = 1$ or in other words sets $A$ and $B$ in the
bipartition are both $O(N)$, we get $ m \equiv O(N)$. For all other values of
$\alpha$, we get $m < O(N^ \alpha)$. As $\alpha$ decreases from $1$, $m$ also
decreases until it becomes $O(1)$ and this occurs for the first time when
$\alpha = 1/(\lambda-1)$. Thus, from (\ref{cab}) and (\ref{E.6})
we get 
$Q_{X^*} \ge O\left(N^{-\frac{\lambda}{\lambda-1}}\right)$ and so 
\begin{equation}
B_T \leq B_e \equiv O\left(N^{\frac{\lambda}{\lambda-1}}\right).  
\label{E.8}
\end{equation}
From (\ref{E.8}) we see that when  $\lambda \to 2$, we get 
the worst possible scaling of $B_e = O(N^2)$,
which can be understood from the fact that  the graph
becomes increasingly star-like, and for such a graph the central node
trivially has $B=O(N^2)$.   On the other hand, when
$\lambda \to \infty$, $B_e \to O(N)$. In this case 
the graph approaches a random regular graph and random
regular graphs are good {\em vertex expanders} \cite{Sarnak}. This implies that
for any bipartition into $A$ and $B$, there exists a constant $\mu$
such that $|c(B)| \ge \mu |A|$. Thus $|c(A)| \ge \frac{\mu}{1 +
\mu} |A|$, so $|c(A)|$ and $|c(B)|$ are linear in 
$|A|$ and hence $B_e = O(N)$.

When $2 < \lambda < 3$, for the networks generated by the configuration model
to be uncorrelated requires that the maximum degree in the
network $K_{max} \sim N^{1/2}$ \cite{Satorras}. Incorporating
this upper cutoff in the arguments made above, we obtain
$Q_{X^*} \ge O(N^{\frac{3}{2}})$ and hence $B_T \le B_e \equiv
O(N^{\frac{3}{2}})$ (same as for $\lambda = 3$ in (\ref{E.8}) ).
From the inset in Fig 3, we see that the scaling the maximal betweenness
incurred by the SP protocol, $B^{SP} \sim N^{1.80}$. This is much worse than
the scaling of $B_e$, and therefore suggests that an SRP for which the maximal
betwenness scales like $B_{e}$ would have a better performance than the SP protocol
from the point of view of congestion. The question arises whether $B_e$ can be
achieved by {\em any}  static routing protocol. We answer this question affirmatively by 
presenting next an SRP for which the scaling of the maximal betwenness is
superior even to the scaling of $B_e$ and therefore significantly better than
the scaling of $B^{SP}$.

  Our derivation of $B_{e}$ suggests that the sparsity is
smallest when obtained from a bipartition where the smaller set is of size of
the order of the maximal degree. This suggests that, topologically, the
betweenness for hubs is high, and using the SP protocol increases this
betweenness since shorter paths largely tend to use hubs. Moreover, using
the SP protocol leaves a large number of alternate paths
unused for routing. Exploiting these observations, we obtain a novel SRP, which we
call the {\em hub avoidance} (HA) protocol, as follows:  (1) Remove $x$ of the
highest degree nodes. The network could now consist of several disconnected
clusters. In every such cluster, assign a routing path for every pair of
nodes using SP.  (2) Place back the removed nodes with their edges. For every
pair of nodes which have not been assigned a routing path in Step 1), assign
one using the SP protocol.  For our simulations we have chosen $x = 0.01 N$, but for optimal
performance the functional dependence of $x$ on $N$ may be different. A detailed theory for this protocol
with these considerations will be presented elsewhere.  Here our primary
purpose of presenting the HA protocol is to indicate that there exists an SRP
for which the scaling of the maximal betweenness not only achieves, but
surpasses the scaling of the topological estimate $B_{e}$, and therefore is a
significant improvement over the SP protocol. This improvement comes from utilizing available alternate paths which, while
not significantly longer than the shortest path,
also considerably alleviate the load on the hubs. The plot in Fig.3 shows the
improvement in performance achieved by our protocol as reflected by the
increase in the position of the congestion threshold and the lowering in the
number of accumulating packets at a given packet creation rate $\gamma$ as
compared to the shortest path protocol. 

Thus, in summary, we identify a bound to communication arising purely due to the
network topology and utilize this to show that there exist better SRPs than
the SP protocol for routing on scale free networks.

We thank Donald Thompson for providing latency data and
J. \v{Z}ivkovi\'{c} for comments. S.S. and H.E.S. were
supported by ONR, E.L. and Z.T. were supported by DOE contract
No. W-7405-ENG-36 and R. Cohen was supported by ISF and DysoNet.

\end{document}